\documentclass[aps,prb,twocolumn,floatfix,showpacs,superscriptaddress]{revtex4}

\usepackage{graphicx}
\usepackage{bbm}
\usepackage{amsmath,amssymb}
\newcommand{\be}{\begin{equation}}
\newcommand{\beq}{\begin{equation}}
\newcommand{\ee}{\end{equation}}
\newcommand{\bea}{\begin{eqnarray}}
\newcommand{\eea}{\end{eqnarray}}
\newcommand{\ba}{\begin{array}}
\newcommand{\ea}{\end{array}}

\renewcommand{\vr} {{\bf r}}
\newcommand{\vj} {{\bf j}}

\begin{document}
\title{Parameter-free density functional for the correlation energy in two dimensions}
\author{E. R{\"a}s{\"a}nen}
\email[Electronic address:\;]{esa.rasanen@jyu.fi}
\affiliation{Nanoscience Center, Department of Physics, 
University of Jyv{\"a}skyl{\"a}, FI-40014 Jyv{\"a}skyl{\"a}, Finland}
\author{S. Pittalis}
\affiliation{Department of Physics and Astronomy, University of Missouri,
Columbia, Missouri 65211, USA}
\author{C. R. Proetto}
\altaffiliation{Present address: 
Centro At{\'o}mico Bariloche and Instituto Balseiro, 8400
S.C. de Bariloche, R{\'i}o Negro, Argentina}
\affiliation{Institut f{\"u}r Theoretische Physik,
Freie Universit{\"a}t Berlin, Arnimallee 14, D-14195 Berlin, Germany}
\affiliation{European Theoretical Spectroscopy Facility (ETSF)}

\date{\today}

\begin{abstract}
Accurate treatment of the electronic correlation in inhomogeneous
electronic systems, combined with the ability to capture the correlation
energy of the homogeneous electron gas, allows to reach high
predictive power in the application of density-functional theory. 
For two-dimensional systems we can achieve  this goal
by generalizing our previous 
approximation [Phys. Rev. B {\bf 79}, 085316 (2009)]
to a parameter-free form, which reproduces the
correlation energy of the homogeneous gas while preserving the
ability to deal with inhomogeneous systems.
The resulting  functional is shown to be very accurate for finite systems
with an arbitrary number of electrons with respect to numerically 
exact reference data.
\end{abstract}


\maketitle

\section{Introduction}
With the present technology, 
electron gas can be confined in various ways to create
nanoscale devices of lower dimensionality. In particular, the
field of two-dimensional (2D) physics has grown rapidly 
alongside the development of electronic (quasi-)2D devices such as
quantum Hall bars and point contacts, and semiconductor quantum 
dots (QDs). 
In the modeling of QDs, the most common and usually
valid approach is to consider a purely 2D Hamiltonian with a standard,
(not screened nor softened) Coulomb interaction and effective
values for the electron mass and dielectric constant, which account for
the surrounding semiconducting material such as GaAs.~\cite{qd}

QDs are studied theoretically in various ways including analytic
methods, exact diagonalization~\cite{rontani} (ED), variants of
quantum Monte Carlo~\cite{harju,pederiva,guclu} (QMC) techniques, Hartree-Fock
methods~\cite{cavaliere}, and density-functional
theory~\cite{dft,note} (DFT). The applicability of DFT crucially depends on
the approximation for the exchange-correlation energy functional. In
spite of recent advances in DFT tailored for strongly correlated 
electrons,~\cite{paola} or in the development of 
2D functionals~\cite{x1,x2,gga,gamma,prev,2dcs,simple,2DBJ} 
beyond the commonly used 2D local spin-density 
approximation~\cite{rajagopal,attaccalite} (LSDA),
there is still a long
path ahead to reach the predictability and efficiency that DFT has in 
quantum chemistry.

In the present work we focus on the correlation energy of
inhomogeneous 2D systems within DFT. 
In particular, we consider a generalization of the
2D functional developed in Ref.~\onlinecite{prev}, which
was based on correlation-hole modeling similar to
that of Becke~\cite{becke,becke_canada} in 3D systems.
In this functional several exact constraints are satisfied, and
the electron spins and currents are incorporated in a natural way 
allowing to deal with spin-polarized and/or current-carrying states.
However, the functional
depends on two constants which enter the estimation of 
the characteristic size of the 
correlation hole (see below), and they
were chosen  {\em semi-empirically} by fitting the
correlation energy to particular {\it finite} systems. Although a 
good performance was obtained, the desired tendency to
work for an arbitrary number of particles was missed.
Here, we show how this limitation can
be overcome by transforming the above-mentioned {\em arbitrary} 
constants to 
{\em non-arbitrary functionals} of the particle density.
This is achieved, as explained in detail below, 
by enforcing the functional
to reproduce the correlation energies of the homogeneous 
2D electron gas (2DEG). 
As a result, not only the correlation
energies of few-electron QDs are reproduced very accurately, broadly
outperforming the standard LSDA approximation, but more importantly
the favorable tendency of increasing accuracy with the number
of electrons is obtained as well. This makes the functional
a predictive tool to calculate correlation energies
in realistic 2D systems.

The rest of the paper is organized as follows.
In Sec.~\ref{I} we briefly review the framework behind
the functional, which is similar to that in Ref.~\onlinecite{prev}.
In Sec.~\ref{II} we determine the coefficients for the
characteristic sizes of the correlation holes by fitting
to the correlation energies in the 2DEG, 
both in the fully-polarized and unpolarized
situations, respectively. In Sec~\ref{III} we apply the resulting expressions 
to QD systems for which accurate reference data is available.
Conclusions are given in Sec.~\ref{IV}.

\section{Modeling the correlation hole}
\label{I}
We start with the formal expression for the correlation energy 
which can be written in
Hartree atomic units (a.u.) as
\be\label{ECH}
E_{c}[\rho_{\uparrow},\rho_{\downarrow}]= \frac{1}{2} \sum_{\sigma\sigma'} \int d\vr_1 \int d\vr_2
\frac{\rho_{\sigma}(\vr_1)}{|\vr_1-\vr_2|}\,h^{\sigma\sigma'}_{c}(\vr_1,\vr_2),
\ee
where
$h^{\sigma\sigma'}_{c}(\vr_1,\vr_2)$ is the correlation-hole function.
In Ref.~\onlinecite{prev} we considered its cylindrical average
$\bar{h}_{c}^{\sigma\sigma'}(\vr=\vr_1,s=|\vr_2-\vr_1|)$ and
constructed a model
satisfying the (i) exact normalization of the spin-dependent correlation-hole functions (sum rule); (ii) correct short-range
behavior for $s\rightarrow 0$, obtainable from the cusp conditions for
the 2D electronic wave function; and (iii) proper decay in the limit
$s\rightarrow\infty$, for which we used a Gaussian approximation.
For the correlation-hole {\em potentials}
\be
\label{chole}
U^{\sigma\sigma'}_c(\vr)= 2 \pi \int_{0}^{\infty} ds \, {\bar h}^{\sigma\sigma'}_{c}(\vr,s) \; .
\ee
we obtained  the expressions
\bea \label{u1}
U^{\sigma\sigma}_c(\vr) & = & \frac{16}{81\pi}\left(8-3\pi\right) D_\sigma(\vr)
z^2_{\sigma\sigma}(\vr) \nonumber \\
& \times & \left[2 z_{\sigma\sigma}(\vr)-3\ln \left( \frac{2}{3}z_{\sigma\sigma}(\vr)+1\right)\right],
\eea
and 
\bea \label{u2}
U^{\sigma{\bar \sigma}}_c(\vr) & = & (2-\pi)\rho_{{\bar \sigma}}(\vr) \nonumber \\
& \times & \left[2
  z_{\sigma{\bar \sigma}}(\vr)-\ln\left(2 z_{\sigma{\bar \sigma}}(\vr) +1\right)  \right]
\eea
for the same- and opposite-spin cases, $\sigma\sigma'=\sigma\sigma$
and $\sigma\sigma'=\sigma{\bar\sigma}$, respectively. 
The spin-pair components of the correlation energy can be calculated from
\be
\label{Ec}
E^{\sigma\sigma'}_{c} = \frac{1}{2} \int d\vr\, \rho_{\sigma}(\vr)
\,U^{\sigma\sigma'}_{c}(\vr) \; ,
\ee
so that the total correlation energy is given by
\be
\label{Ectot}
E_c [\rho_{\uparrow},\rho_{\downarrow}] =E^{\uparrow\uparrow}_{c} + E^{\downarrow\downarrow}_{c} + 
2 E^{\uparrow\downarrow}_{c}
\ee
with the condition $E^{\uparrow\downarrow}_{c}=E^{\downarrow\uparrow}_{c}$.

Let us next examine the ingredients of Eqs.~(\ref{u1}) and (\ref{u2}).
First, in Eq.~(\ref{u1}) we have
\be
D_{\sigma}:= \frac{1}{2} \left( \tau_\sigma - \frac{1}{4} \frac{\left( \nabla \rho_\sigma 
\right)^2}{\rho_\sigma} - \frac{\vj^2_{p,\sigma}}{\rho_\sigma} \right)
\label{D}
\ee
containing two quantities that depend on the occupied Kohn-Sham
orbitals: $\tau_\sigma=\sum_{k=1}^{N_\sigma} |\nabla\psi_{k,\sigma}|^2$
is (double) the kinetic-energy density,
and $\vj_{p,\sigma}=\frac{1}{2i}\sum_{k=1}^{N_\sigma} \left[
  \psi^*_{k,\sigma} \left(\nabla \psi_{k,\sigma}\right) - \left(\nabla \psi^*_{k,\sigma}\right) \psi_{k,\sigma} \right]$
is the spin-dependent paramagnetic current density. 
It can be easily seen that $D_\sigma(\vr)$ vanishes for all 
single-particle ($N=1$) systems. Hence, $E^{\sigma\sigma'}_{c}$
vanishes as well and the functional is correctly 
self-interaction free for $N=1$ (in contrast with, e.g., the LSDA).

As we have mentioned above, important quantities 
in Eqs.~(\ref{u1}) and (\ref{u2}) are the
characteristic sizes of the correlation holes $z_{\sigma\sigma'}(\vr)$.
They are assumed to be
proportional to the sizes of the corresponding exchange holes, i.e.,
\bea \label{z1}
z_{\sigma\sigma}(\vr) & := & 2 c_{\sigma\sigma}
|U_x^{\sigma}(\vr)|^{-1}, \\
z_{\sigma{\bar \sigma}}(\vr) & := & c_{\sigma{\bar \sigma}} \left[ |U_x^{\sigma}(\vr)|^{-1}+ |U_x^{{\bar \sigma}}(\vr)|^{-1}\right],
\label{z2}
\eea
where $U_x^{\sigma}$ is the exchange-hole potential~\cite{x1,becke_roussel}
for spin $\sigma$. The idea behind the assumption is the following:
the smaller the Fermi hole around each electron is, the more tightly
the electrons are screened, and therefore they are expected to be
correlated much less. This coarse picture of the real situation has
been found to work very well in 
practice.~\cite{prev,becke,becke_canada,becke_current}

\section{Determination of coefficients $c_{\sigma\sigma'}$}
\label{II}

In Ref.~\onlinecite{prev} the coefficients 
$c_{\sigma\sigma}$ and $c_{\sigma{\bar \sigma}}$ 
were {\em constants} determined by 
fitting the total correlation energy
of a set of parabolic QDs. In the present work,
these spin-dependent coefficients
are expressed as functionals of the particle density.
We achieve this goal in two steps. First, 
we determine the values of  $c_{\sigma\sigma'}$
that yield the {\em exact} correlation energy 
density~\cite{attaccalite} of the polarized (unpolarized) 2DEG,
denoted below as $\epsilon_{c}^{\rm 2DEG}[r_s,\zeta=1]$
($\epsilon_{c}^{\rm 2DEG}[r_s,\zeta=0]$), where $\zeta$ is 
the spin polarization. This allows us to write
\be
c_{\sigma\sigma'}\rightarrow c_{\sigma\sigma'}[r_s].
\ee
where  $r_s=1/\sqrt{\pi\rho}$ and  
$\rho=\rho_\uparrow+\rho_\downarrow$ is the total density of the 2DEG.
Then as a second step, when the functional is applied to an 
inhomogeneous system, we express  $r_s$  in terms of the {\em local} 
particle density, i.e., for each point in space.
This is nothing else but a local-density approximation for the
coefficients $c_{\sigma\sigma'}$.  Otherwise, the correlation functional
has the form given in Eqs.~(\ref{chole}-\ref{Ec}). Hence, it
still satisfies all the exact constraints listed above and the 
overall expression is of the form of a current-dependent 
meta-generalized-gradient approximation (meta-GGA).

\subsection{Fully-polarized case}

\begin{figure}
\includegraphics[width=0.80\columnwidth]{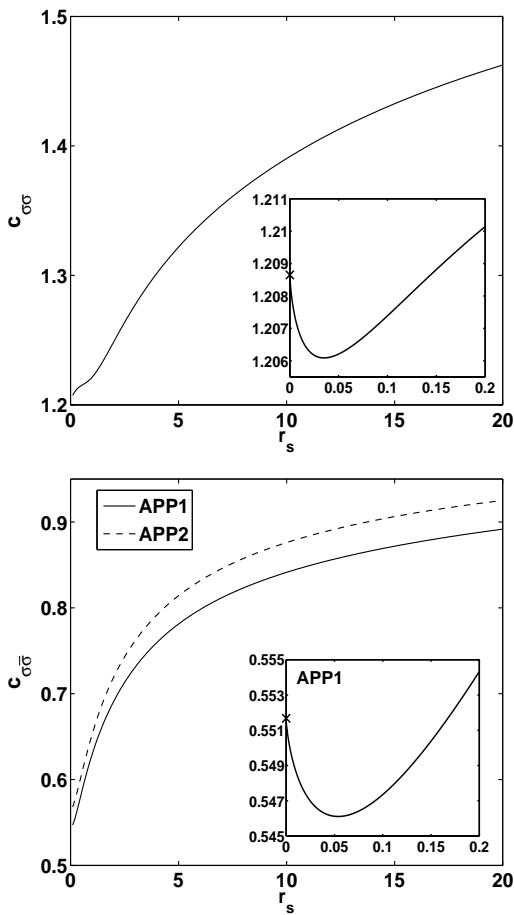}
\caption{Upper panel: Optimal coefficients $c_{\sigma\sigma}$ for
locally reproducing the correlation energy densities of the
homogeneous two-dimensional electron gas. The inset show the
detailed behavior in the high-density limit $r_s\rightarrow 0$.
Lower panel: Same for $c_{\sigma\bar{\sigma}}$. Solid and dashed
lines correspond to two difference approximations for
the polarized component of the correlation energy in the
unpolarized electron gas (see text). The inset corresponds to
the high-density result in the first approximation (APP1).
}
\label{fig1}
\end{figure}

Now we demonstrate in detail the procedure determining
 $c_{\sigma\sigma'}\rightarrow c_{\sigma\sigma'}[r_s]$.
First we focus on the fully-polarized case with 
$\sigma\sigma'=\sigma\sigma$, where $\sigma=\uparrow$ 
or $\sigma=\downarrow$, and thus $\zeta=1$. 
In this case $\rho = \rho_\uparrow$ {\em or} $\rho = \rho_\downarrow$. 
Comparing Eq.~(\ref{Ec}) with the LSDA expression,
\be
\label{lsda}
E_{c}^{\rm LSDA}=\int d\vr\, \rho(\vr)\,\epsilon_{c}^{\rm 2DEG}[r_s,\zeta=1]
\ee
directly yields $U_c^{\sigma\sigma}=2\,\epsilon_{c}^{\rm 2DEG}[r_s,\zeta=1]$.
Similarly, for the exchange-hole potential in 
Eq.~(\ref{z2}) (see Ref.~\onlinecite{prev} for its definition) we
find $U_x^{\sigma}=2\,\epsilon_{x}^{\rm 2DEG}[r_s,\zeta=1]=-16/(3\pi
r_s)$.
From Eq.~(\ref{z1}) we get 
$z_{\sigma\sigma}=2\,c_{\sigma\sigma}|U_x^{\sigma}|^{-1}=3\pi
c_{\sigma\sigma}r_s/8$, 
and Eq.~(\ref{D}) for the 2DEG gives 
$D_\sigma=\pi^{-1}r_s^{-4}$.
Collecting these
results to Eq.~(\ref{u1}) leads to
\begin{multline}
\label{c_formula1}
\epsilon_{c}^{\rm 2DEG}[r_s,\zeta=1]  = \\
\frac{(8-3\pi)}{24}\frac{c^2_{\sigma\sigma}}{r_s^2}\Bigg[\frac{\pi}{4}c_{\sigma\sigma}\,r_s
-  \ln\left(\frac{\pi}{4}\,c_{\sigma\sigma}\,r_s+1\right)\Bigg].
\end{multline}
This expression can be solved numerically for
$c_{\sigma\sigma}[r_s]$ by using the parametrized QMC result
for the correlation energy of the 2DEG.~\cite{attaccalite}
The result is shown in the upper panel of Fig.~\ref{fig1}
for realistic densities ($0<r_s<20$). The data could be easily
tabulated or parametrized for convenient use of the functional.
For this range of densities we may use approximate 
parametrization of the form
\be
\label{c_par1}
c_{\sigma\sigma}[r_s]\approx \alpha\,\log(r_s)+\beta\,r_s^{\gamma}
\ee
with $\alpha=-0.1415\,1$, $\beta=1.226\,1$, and $\gamma=0.144\,99$.
We remind that when using the correlation functional the
coefficient $c_{\sigma\sigma}[r_s](\vr)$ is calculated locally 
with the local density $r_s(\vr)$.

Let us examine in detail the high-density limit, $r_s\rightarrow 0$,
for the same-spin coefficient.
Expansion of the logarithm in Eq.~(\ref{c_formula1}) leads to
\bea
\label{css}
c_{\sigma\sigma}[r_s\rightarrow 0] & = &\left\{\frac{768}{\pi^2
  (8-3\pi)}\epsilon_{c}^{\rm 2DEG}[r_s\rightarrow
0,\zeta=1]\right\}^{1/4} \nonumber \\
& \approx & 1.2087,
\eea
where we have used the known limit for the 2DEG, 
$\epsilon_{c}^{\rm 2DEG}[r_s\rightarrow
0,\zeta=1]\approx -0.039\,075\,0$ (Ref.~\onlinecite{attaccalite}).
This result coincides with the numerical one in 
Fig.~\ref{fig1} (see the upper inset). We point out that
the ``u-shape'' at $r_s\lesssim 0.05$ is not due to numerical
inaccuracies in the present work or in the QMC parametrization,
but it arises from the exact properties of the 2DEG correlation
energy in the high-density limit.~\cite{rajagopal}

In the low-density limit, $r_s \rightarrow \infty$, we obtain from 
Eq.~(\ref{c_formula1}) the leading contribution as
\bea \label{cssinfty}
c_{\sigma \sigma}[r_s \rightarrow \infty] & \simeq &
\left\{ \frac{96\,r_s}{\pi(8-3\pi)} \epsilon_{c}^{\rm 2DEG}[r_s \rightarrow \infty,\zeta = 1] \right\}^{1/3} \nonumber \\
& \approx & 1.68,
\eea
where we have used the known result for the 2DEG, 
$\epsilon_{c}^{\rm 2DEG}[r_s \rightarrow \infty,\zeta = 1] \rightarrow -0.222/r_s$ (Ref.~\onlinecite{attaccalite}).
From the upper panel of Fig.~\ref{fig1} it is 
evident that this extreme low-density 
limit has been not reached yet for $r_s \simeq 20$. 
However, our numerical values at larger $r_s$ agree
very well with the exact limit.

\subsection{Unpolarized case}

The coefficient $c_{\sigma{\bar \sigma}}[r_s]$ can be determined in a
similar fashion by employing the results for the 
correlation energy density of the {\em unpolarized} 2DEG,
$\epsilon_{c}^{\rm 2DEG}[r_s,\zeta=0]$. It should be noted that
also the same-spin component $U_c^{\sigma\sigma}$ 
depending on the spin density $\rho_\sigma$
is needed in the calculation. 
Now, the fitting to the 2DEG raises the following 
complication: the parallel-spin contribution of
$\epsilon_{c}^{\rm 2DEG}[r_s,\zeta=0]$, which we here
mark as $\epsilon_{c,{\sigma\sigma}}^{\rm 2DEG}[r_s,\zeta=0]$,
is nontrivial; in particular, it is important to note that this contribution 
is not the same as $\epsilon_{c}^{\rm 2DEG}[r_s,\zeta=1]$ 
upon density scaling. In principle, a fully
spin-resolved expression is accessible by employing 
the QMC results for the 2DEG correlation potential energy 
and using the virial theorem.~\cite{paola_g}
For the sake of simplicity, however, we apply here two 
different approximations for the parallel-spin component.
In the first approximation (APP1) 
we set $\epsilon_{c,{\sigma\sigma}}^{\rm 2DEG}[r_s,\zeta=0]
 \approx  \epsilon_{c}^{\rm 2DEG}[\sqrt{2}\,r_s,\zeta=1]$,
which is similar to the form of Stoll {\em et al.}~\cite{stoll}
in 3D electron gas. In fact, this approximation corresponds to the density 
scaling mentioned above. The second possible approximation 
(APP2) is given by $\epsilon_{c,{\sigma\sigma}}^{\rm 2DEG}[r_s,\zeta=0]
 \approx \epsilon_{c}^{\rm 2DEG}[r_s,\zeta=1]/\sqrt{2}$,
which is similar to the 3D version of Perdew and Wang.~\cite{pw}
Even though the first approximation is {\em exact} in the 
limit $r_s\rightarrow 0$, the latter type of approximation has 
been shown to be more accurate~\cite{gorigiorgi}
in 3D, and we may thus expect similar tendency in 2D as well. 
It should be noted, however, that deviations from numerically 
exact results have been reported in 3D for both approximations,
especially at large $r_s$ (Ref.~\onlinecite{paolaperdew}).

Going back to the determination of $c_{\sigma{\bar \sigma}}[r_s]$,
we consider spin-densities $\rho_\uparrow=\rho_{\downarrow}=\rho/2$
and set a condition $U_c^{\sigma\sigma}=2\,\epsilon_{c,\sigma\sigma}^{\rm 2DEG}[r_s,\zeta=0]$
with the approximations APP1 and APP2 for the r.h.s. as specified above.
From Eqs.~(\ref{u2})-(\ref{Ectot})
the formula to be solved for $c_{\sigma{\bar \sigma}}[r_s]$ is given by
\bea \label{csz}
&& \epsilon_{c,\sigma\sigma}^{\rm 2DEG}[r_s,\zeta=0] - \epsilon_{c}^{\rm 2DEG}[r_s,\zeta=0] = \frac{(\pi-2)}{4 \pi r_s^2} \nonumber \\ 
&&  
\times \left[\frac{3\pi\sqrt{2}}{4} r_s c_{\sigma{\bar \sigma}} -
\ln\left(1 + \frac{3\pi\sqrt{2}}{4} r_s c_{\sigma {\bar \sigma}} \right)\right].
\eea
The result is shown in the lower part of Fig.~\ref{fig1} for both APP1 (solid line) 
and APP2 (dashed line). Both curves for this
density range ($0<r_s<20$) can be approximated 
with a satisfactory accuracy by a simple parametrized formula
\be\label{c_par2}
c_{\sigma{\bar \sigma}}[r_s] \approx \delta\,r_s^{\xi}
\ee
with $\delta=0.653\,58$ and $\xi=0.116\,91$ for APP1
 and $\delta=0.663\,25$ and $\xi=0.123\,96$ for APP2.

Again, let us consider the high- and low-density limits for 
the obtained expressions.
In the high-density limit we find a semi-analytic expression
\begin{multline} \label{csz0}
c_{\sigma{\bar \sigma}}[r_s\rightarrow 0] =
\frac{8}{3\sqrt{\pi(\pi-2)}} \, \bigg\{\epsilon_{c,\sigma\sigma}^{\rm 2DEG}[r_s\rightarrow 0,\zeta=0] \\
                                          -\epsilon_{c}^{\rm 2DEG}[r_s\rightarrow 0,\zeta=0] \bigg\}^{1/2}.
\end{multline}
Using the known limit $\epsilon_{c}^{\rm 2DEG}[r_s\rightarrow
0,\zeta=0]\approx -0.192\,500$ (Ref.~\onlinecite{attaccalite}) leads
 to $c^{\rm APP1}_{\sigma{\bar \sigma}}[r_s\rightarrow 0]\approx 0.55168$,
which corresponds to the exact result in this limit as mentioned above,
and $c^{\rm APP2}_{\sigma{\bar \sigma}}[r_s\rightarrow 0]\approx 0.57175$.
Both values coincide with the numerical results in Fig.~\ref{fig1} 
(see the lower panel for a detailed view on APP1). 

In the low-density limit Eq.~(\ref{csz}) yields
\begin{multline} \label{cszinfty}
 c_{\sigma \bar{\sigma}}[r_s \rightarrow \infty] =
\frac{16\,r_s}{3 \sqrt{2} (\pi-2)} \bigg\{ \epsilon_{c,\sigma\sigma}^{\rm 2DEG}[r_s \rightarrow \infty,\zeta = 0] \\
- \epsilon_{c}^{\rm 2DEG}[r_s \rightarrow \infty,\zeta = 0]\bigg\}
\end{multline}
From the knowledge that 
$\epsilon_{c}^{\rm 2DEG}[r_s \rightarrow \infty,\zeta = 0] \rightarrow -0.470/r_s$ 
(Ref.~\onlinecite{attaccalite}), 
we obtain $c_{\sigma \bar{\sigma}}[r_s \rightarrow \infty] \simeq 1.03$. 
Note that the limit value is the {\em same} for APP1 and APP2, which 
is obvious due to the low-density decay of the correlation energy 
as $\propto 1/r_s$ 
in the fully-polarized 2DEG (see the previous section).
The congruence in the limit value between APP1 and APP2
is not evident in the density range ($r_s\leq 20$)
shown in Fig.~\ref{fig1}. In fact, the decay towards the
same limit can be seen only at very high $r_s$ as we have confirmed
numerically.

It is reassuring to note that, to the leading order, both high- and
low-density limits of our model correlation energies obtained through 
Eqs.~(\ref{c_formula1}) and (\ref{csz})
are in agreement with the corresponding exact expansions regarding their
functional dependence with respect to the density parameter $r_s$. More
specifically, $\epsilon_c^{\rm 2DEG}[r_s \rightarrow 0, \zeta = 0]$ and
$\epsilon_c^{\rm 2DEG}[r_s \rightarrow 0, \zeta = 1]$ attain a finite ($r_s$
independent) value in the high-density limit, while $\epsilon_c^{\rm 2DEG}[r_s
\rightarrow \infty, \zeta = 0]$ and $\epsilon_c^{\rm 2DEG}[r_s \rightarrow
\infty, \zeta = 1]$ decay as $r_s^{-1}$ in the low-density limit.

\begin{figure}
\includegraphics[width=0.80\columnwidth]{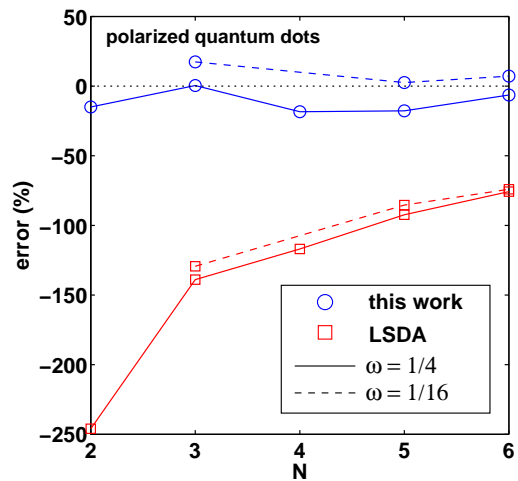}
\caption{(color online). Relative errors produced by the
present functional (circles) and the local spin-density approximation
(squares) in correlation
energies of spin-polarized quantum dots with respect to numerically 
accurate results in Ref.~\onlinecite{rontani} (see text).}
\label{fig2}
\end{figure}

\section{Testing on inhomogeneous systems}
\label{III}

Next we test the functional, i.e., Eqs.~(\ref{u1}-\ref{z2}) with
the coefficients $c_{\sigma\sigma}$ and $c_{\sigma{\bar \sigma}}$ taken 
from Eqs.~(\ref{c_par1}) and (\ref{c_par2}), respectively,
for a set of 2D parabolic quantum
dots, where the external confining potential is given by 
$v(r)=\omega^2 r^2/2$ with $\omega$ being the confinement strength.
The correlation energies obtained from Eq.~(\ref{Ec}) are
compared to the reference results $E_c^{\rm ref}=E_{\rm tot}^{\rm ref}-E_{\rm tot}^{\rm EXX}$,
where $E_{\rm tot}^{\rm ref}$ is the analytic,~\cite{taut} ED,~\cite{rontani}
or QMC,~\cite{pederiva} total energy, and $E_{\rm tot}^{\rm EXX}$ is
the exact-exchange (EXX) total energy 
obtained in the Krieger-Li-Iafrate approximation~\cite{kli}
with the {\tt octopus} code.~\cite{octopus}
The EXX result is used as input for our functional,
including $U_x^\sigma$ as the exact
exchange-hole potential. 

\subsection{Fully-polarized quantum dots}

Figure~\ref{fig2}
shows the relative errors in the correlation energies for
fully-polarized QDs of $N=2\ldots 6$ electrons and
confinement strengths $\omega=1/4$ (solid lines) and $\omega=1/16$
(dashed lines). The errors produced by the present functional 
and by the LSDA with respect to the ED results
in Ref.~\onlinecite{rontani} are presented by circles and squares, respectively.
The present functional clearly outperforms the LSDA by reducing the
error in the correlation energy by one order of magnitude on the
average (mean absolute errors $120\,\%$ and $11\,\%$, respectively).
The performance of the present functional is stable regardless of
$N$ and $\omega$, whereas the LSDA gains accuracy very slowly as
a function of $N$.
Since our functional coincides with the LSDA in the 2DEG limit by
construction, it can be expected that the error of the present
functional diminishes further at larger $N$. This desired tendency is 
clearly missed if a fixed value for $c_{\sigma\sigma}$ is used.~\cite{prev}

\subsection{Unpolarized quantum dots}

In Table~\ref{table}
\begin{table}
  \caption{\label{table} Comparison of the correlation energies for
unpolarized parabolic quantum dots. The reference results have been
obtained from numerically accurate data for the total energy (see text). 
APP1 and APP2 correspond to our functional with two different approximations
for the parallel-spin component, respectively (see text).
The last row shows the mean absolute error (in percentage), 
$\Delta=\left<|E_c-E_c^{\rm  ref}|/|E_c^{\rm ref}|\right>$. 
}
  \begin{tabular*}{\columnwidth}{@{\extracolsep{\fill}} c c c c c c c}
  \hline
  \hline
  $N$ & $\omega$ & $E_{\rm tot}^{\rm ref}$ & $E_{\rm c}^{\rm ref}$ & $E_c^{\rm APP1}$ & $E_c^{\rm APP2}$ & $E_c^{\rm LSDA}$ \\
  \hline
2  & 1    & $3^*$                   &-0.162  &  -0.144 & -0.147 & -0.199 \\
2  & 1/4  & $0.9324^\dagger$         &-0.114  &  -0.098 & -0.101 & -0.139 \\
2  & 1/6  & $2/3^*$                 &-0.102  &  -0.085 & -0.087 & -0.122 \\
2  & 1/16 & $0.3031^\dagger$         &-0.070  &  -0.057 & -0.059 & -0.085 \\
2  & 1/36 & $0.1607^\dagger$         &-0.049  &  -0.040 & -0.041 & -0.061 \\
6  & $1/1.89^2$ & $7.6001^\ddagger$  & -0.421 &  -0.396 & -0.405 & -0.473 \\
6  & 1/4  &  $6.995^\dagger$         &-0.396  &  -0.381 & -0.390 & -0.457 \\
6  & 1/16 & $2.528^\dagger$          &-0.250  &  -0.221 & -0.228 & -0.279 \\
12 & $1/1.89^2$ & $25.636^\ddagger$  & -0.917 &  -0.895 & -0.915 & -1.000 \\
   \hline
$\Delta$ & & & & $11\,\%$ & $9\,\%$ & $18\,\%$\\
\hline
  \hline
  \end{tabular*}
  \begin{flushleft}
\vspace{-2mm}
    $^*$   Analytic solution by Taut from Ref.~\onlinecite{taut}.
  $^\dagger$  CI data from Ref.~\onlinecite{rontani}.
  $^\ddagger$ Diffusion QMC data from Ref.~\onlinecite{pederiva}.
  \end{flushleft} 
  \end{table}
we compare the correlation energies for a set of unpolarized
QDs. Also in this case the present functional, with both approximations 
APP1 and APP2 for the parallel-spin component, is 
more accurate than the LSDA.
As expected, LSDA becomes again more accurate with $N$, but
this tendency can be found also in the present
functional, where it is in fact significantly stronger than in the LSDA.
For example, for $N=2$, $N=6$, and $N=12$ with roughly the same
$\omega$, the LSDA yields relative errors of
$-22\,\%$, $-12\,\%$, and $-9\,\%$ in the correlation energies, respectively.
When using our functional with APP1 the corresponding errors
are $14\,\%$, $6\,\%$, and $2\,\%$,
respectively, and within APP2 the errors reduce further 
to $11\,\%$, $1.5\,\%$, and $0.2\,\%$.
As can be seen in Ref.~\onlinecite{prev},
the functional having a fixed value for $c_{\sigma{\bar\sigma}}$
is far from this accuracy.
The better performance of APP2 in comparison with APP1 is in
line with the results for the 3D electron gas, where the
approximation of Perdew and Wang~\cite{pw} (similar to APP2)
is more accurate than the one by Stoll {\em et al.}~\cite{stoll}
(similar to APP1).

Before concluding, it is natural to ask if the exchange-hole potential 
$U_x^{\sigma}(\vr)$ may be treated approximately to reduce the burden of the
EXX calculation without losing the accuracy.
Possible choices may be the LSDA expression, or the 
GGA and meta-GGA as 
given in Refs.~\onlinecite{gga,gamma,x1,x2}. 
Also the goal is to carry out fully 
self-consistent calculations and preferably for a larger 
variety of 2D systems, e.g., for quantum-ring structures.
Here we have focused solely on parabolic QDs representing,
through comparison with experimental data,~\cite{impurity,spindroplet,max} 
a valid approximation for both 
vertical and lateral semiconductor QD devices.
The above outlined tasks clearly
deserve extended future investigations.

\section{Conclusion}
\label{IV}

We have developed a spin-dependent parameter-free density functional to 
calculate the correlation energies in two-dimensional electron
systems. The functional has been constructed through physically 
reasonable modeling of the angular-averaged, 
spin-resolved correlation hole functions.
The key extension to previous works is
enforcing the functional to reproduce the correlation energies
of the homogeneous two-dimensional electron gas. 
We have shown that this is possible by transforming the 
coefficients -- involved in the estimation of the characteristic 
sizes of the correlation holes --
to functionals of the electron density. As a result,
we are able to find very accurate correlation energies for quantum 
dots with varying confinement strength and number of electrons.
Most importantly, the significant error reduction as a function of
the number of electrons 
makes the present functional a predictive method to obtain
correlation energies of systems which are beyond the capabilities
of exact-diagonalization and quantum Monte Carlo techniques. 

\begin{acknowledgments}
This work was supported by the Academy of Finland and 
the EU's Sixth Framework Programme through the ETSF e-I3.
C. R. P. was supported by the European
Community through a Marie Curie IIF (MIF1-CT-2006-040222) 
and CONICET of Argentina through PIP 5254. 
S. Pittalis acknowledges support by DOE grant DE-FG02-05ER46203.
\end{acknowledgments}


\begin{thebibliography}{ll}

\bibitem{qd}
  For a review, see, e.g., L. P. Kouwenhoven, D. G. Austing, and S. Tarucha, Rep.
  Prog. Phys. {\bf 64} (2001) 701; S. M. Reimann and M. Manninen, Rev. Mod.
  Phys. {\bf 74} (2002) 1283.

\bibitem{rontani}
M. Rontani, C. Cavazzoni, D. Bellucci, and G. Goldoni,
J. Chem. Phys. {\bf 124}, 124102 (2006).

\bibitem{harju}
For a review, see, A. Harju, J. Low Temp. Phys. {\bf 140}, 181 (2005).

\bibitem{pederiva}
F. Pederiva, C. J. Umrigar, and E. Lipparini,
Phys. Rev. B {\bf 62}, 8120 (2000); {\em ibid} {\bf 68}, 089901 (2003).

\bibitem{guclu} A. D. G{\"u}\c{c}l{\"u}, J.-S. Wang, and H. Guo,
Phys. Rev. B {\bf 68}, 035304 (2003).

\bibitem{cavaliere}
See, e.g., U. De Giovannini, F. Cavaliere, R. Cenni, M. Sassetti, and B. Kramer, 
Phys. Rev. B {\bf 77}, 035325 (2008).

\bibitem{dft} For reviews about DFT, see, e.g., R. M. Dreizler and E. K. U. Gross,
{\it Density functional theory} (Springer, Berlin, 1990);
U. von Barth, Phys. Scr. {\bf T109}, 9 (2004);
J. P. Perdew and S. Kurth, in A Primer
in Density Functional Theory (Springer, Berlin, 2003).

\bibitem{note} For applications of DFT to quantum dots,
see, e.g., Refs. [9-16] and references therein.

\bibitem{paola}
 P. Gori-Giorgi, M. Seidl, and G. Vignale,
Phys. Rev. Lett. {\bf 103} 166402 (2009).

\bibitem{x1} S. Pittalis, E. R{\"a}s{\"a}nen,
N. Helbig, and E. K. U. Gross, Phys. Rev. B {\bf 76}, 235314 (2007).

\bibitem{x2}
E. R\"as\"anen, S. Pittalis, C. R. Proetto, and E. K. U. Gross,
Phys. Rev. B {\bf 79}, 121305(R) (2009).

\bibitem{gga}
S. Pittalis, E. R\"as\"anen, J. G. Vilhena, M. A. L. Marques,
Phys. Rev. A {\bf 79}, 012503 (2009).

\bibitem{gamma}
S. Pittalis, E. R\"as\"anen,
and E. K. U. Gross, Phys. Rev. A {\bf 80}, 032515 (2009).

\bibitem{prev}
S. Pittalis, E. R\"as\"anen, C. R. Proetto, and E. K. U. Gross,
Phys. Rev. B {\bf 79}, 085316 (2009).

\bibitem{2dcs}
S. Pittalis, E. R{\"a}s{\"a}nen, and M.\,A.\,L. Marques,
Phys. Rev. B {\bf 78}, 195322 (2008).

\bibitem{simple} 
S. Pittalis and E. R{\"a}s{\"a}nen,
Phys. Rev. B {\bf 80}, 165112 (2009).

\bibitem{2DBJ} S. Pittalis, E. R\"as\"anen, and C. R. Proetto,
Phys. Rev. B {\bf 81}, 115108 (2010).

\bibitem{rajagopal}
A. K. Rajagopal and J. C. Kimball,
Phys. Rev. B {\bf 15}, 2819 (1977).

\bibitem{attaccalite}
C. Attaccalite, S. Moroni, P. Gori-Giorgi, and
G. B. Bachelet, Phys. Rev. Lett. {\bf 88}, 256601 (2002).

\bibitem{becke} A. D. Becke, J. Chem. Phys. {\bf 88}, 1053 (1988).

\bibitem{becke_canada} A. D. Becke, Can. J. Chem. {\bf 74}, 995 (1996).

\bibitem{becke_roussel}
A. D. Becke and M. R. Roussel, Phys. Rev. A {\bf 39}, 3761 (1989).

\bibitem{becke_current}
A. D. Becke, J. Chem. Phys. {\bf 117}, 6935 (2002).

\bibitem{paola_g} P. Gori-Giorgi, S. Moroni, and G. B. Bachelet,
Phys. Rev. B {\bf 70}, 115102 (2004).

\bibitem{stoll} H. Stoll, C. M. E. Pavlidou, and H. Preuss, 
Theor. Chim. Acta {\bf 49}, 143 (1978).

\bibitem{pw} J. P. Perdew and Y. Wang, Phys. Rev. B {\bf 46}, 12 947 (1992).

\bibitem{gorigiorgi} P. Gori-Giorgi, F. Sacchetti, and G. B. Bachelet,
Phys. Rev. B {\bf 61}, 7353 (2000). 

\bibitem{paolaperdew} P. Gori-Giorgi and J. P. Perdew,
Phys. Rev. B {\bf 69}, 041103(R) (2004).

\bibitem{taut}
M. Taut, J. Phys. A {\bf 27}, 1045 (1994).

\bibitem{kli} J. B. Krieger, Y. Li, and G. J. Iafrate, Phys. Rev. A
  {\bf 46}, 5453 (1992).

\bibitem{octopus} M. A. L. Marques, A. Castro, G. F. Bertsch and A. Rubio,
  Comp. Phys. Comm. \textbf{151}, 60 (2003);
A. Castro, H. Appel, M. Oliveira, C. A. Rozzi, X. Andrade,
F. Lorenzen, M. A. L. Marques, E. K. U. Gross, and A. Rubio,
Phys. Stat. Sol. (b) {\bf 243}, 2465 (2006).

\bibitem{impurity}  E. R{\"a}s{\"a}nen, J. K{\"o}nemann, 
R. J. Haug, M. J. Puska, and R. M. Nieminen
Phys. Rev. B {\bf 70}, 115308 (2004).

\bibitem{spindroplet} E. R{\"a}s{\"a}nen, H. Saarikoski, 
A. Harju, M. Ciorga, and A. S. Sachrajda
Phys. Rev. B {\bf 77}, 041302(R) (2008).

\bibitem{max} M. Rogge, E. R{\"a}s{\"a}nen, and R. J. Haug,
submitted, arXiv:1001.5395 (2010).


\end{thebibliography}
\end{document}